\documentclass[aps,prl,superscriptaddress,floatfix,twocolumn,longbibliography, nobalancelastpage]{revtex4-2}
\usepackage{amssymb,amsmath,amsfonts}
\usepackage{graphicx}
\usepackage{theorem}
\usepackage{dcolumn}
\usepackage{braket}
\usepackage{bm}
\usepackage{setspace}
\usepackage{eurosym}
\usepackage{amsfonts,amsmath, mathtools}
\usepackage{mathrsfs}
\usepackage{amssymb}
\usepackage{graphicx}
\usepackage{bm}
\usepackage{siunitx}
\usepackage{orcidlink}
\usepackage{enumitem}
\usepackage{comment}
\usepackage{upgreek}

\begin{document}

\title{Substrate Screening for Compact Electron-on-Neon Multiqubit Architectures}
\author{Toshiaki Kanai~\orcidlink{0000-0002-9446-2497}}
\email{Email: kanai@wustl.edu}
\affiliation{Department of Physics, Washington University, St.\,Louis, Missouri 63130, USA}

\author{Chuanwei Zhang}
\email{Email: chuanwei.zhang@wustl.edu}
\affiliation{Department of Physics, Washington University, St.\,Louis, Missouri 63130, USA}
\affiliation{Center for Quantum Leaps, Washington University, St. Louis, Missouri 63130, USA}

\begin{abstract}
Scaling up electron-on-neon devices requires controllable coupling and suppressed crosstalk. We quantify substrate screening of Coulomb interactions and exchange couplings on thin solid neon films.
Metallic screening permits appreciable exchange between nearby qubits while suppressing coupling to more distant electrons.
Comparison with experiment provides model-dependent constraints on trapping configurations.
These results motivate a compact multiqubit architecture that combines the existing superconducting resonator platform with metallic screening and local gate control.
\end{abstract}

\maketitle
Electrons floating above a solid neon surface (e-Ne) provide a promising platform for quantum information processing, in which coherent control and dispersive readout of a single e--Ne qubit have been demonstrated in a circuit quantum electrodynamics architecture~\cite{zhou_single_2022}.
Subsequent experiments~\cite{zhou_electron_2024} achieved relaxation and coherence times on the order of $0.1~\mathrm{ms}$ with an average single-qubit gate fidelity of 99.97\%.
Despite these advances, two central challenges remain.
First, the microscopic origin of lateral electron confinement in existing devices remains unresolved.
Proposed mechanisms involve surface topography and neon-film thickness variations~\cite{kanai_single-electron_2024, kanai_electron_2026, surse_deterministic_2026}, with experiments also indicating a role for substrates~\cite{zheng_surface_2025}.
Second, achieving high-fidelity two-qubit gates remains a major obstacle to scaling.
Both direct Coulomb coupling between electrons~\cite{li_coherent_2025} and indirect coupling through a shared microwave resonator~\cite{zhou_electron_2024, li_solid_2026} are being investigated.
Although coherent two-qubit operations have been demonstrated through direct coupling, the reported microwave-driven operation times are comparable to the relaxation and coherence times of the shorter-lived qubit~\cite{li_coherent_2025}, implying substantial decoherence-induced gate errors.
Improving the ratio of coherence time to gate duration is therefore essential for fault-tolerant operation.

Evaluating the inter-electron Coulomb interaction is critical for scaling the qubit architecture due to its relevance to two-qubit gates and crosstalk.
For electrons above the thin neon films used in e--Ne devices, the underlying dielectric or metallic substrate screens their mutual Coulomb interaction.
Consequently, the coupling cannot generally be inferred from the bulk-neon dielectric response alone.
Quantifying substrate screening is therefore essential for determining the exchange and conditional couplings.
Because inter-electron couplings depend on the electron wavefunctions and their electrostatic environment, measurements of these couplings may provide additional constraints on the trapping configurations realized in experiments.

In this Letter, we quantify substrate screening of Coulomb interactions and the resulting exchange coupling between e--Ne qubits using electrostatic Green functions and self-consistent Hartree calculations.
We show that metallic screening permits appreciable exchange at short separations while suppressing interactions with more distant electrons.
Comparison with reported two-qubit measurements highlights a puzzle: within the assumed model, closely spaced electrons above dielectric regions generally exhibit exchange couplings larger than the measured value, except for a restricted range of dipole orientations.
This comparison provides model-dependent constraints on electron locations and trapping configurations, offering clues to the unresolved microscopic origin of lateral confinement. These results motivate a compact multiqubit architecture that retains the existing superconducting resonator platform while modifying the trapping-region layout and incorporating local gate control.

\begin{figure}[b]
  \centering
    \includegraphics[width=0.92\linewidth]{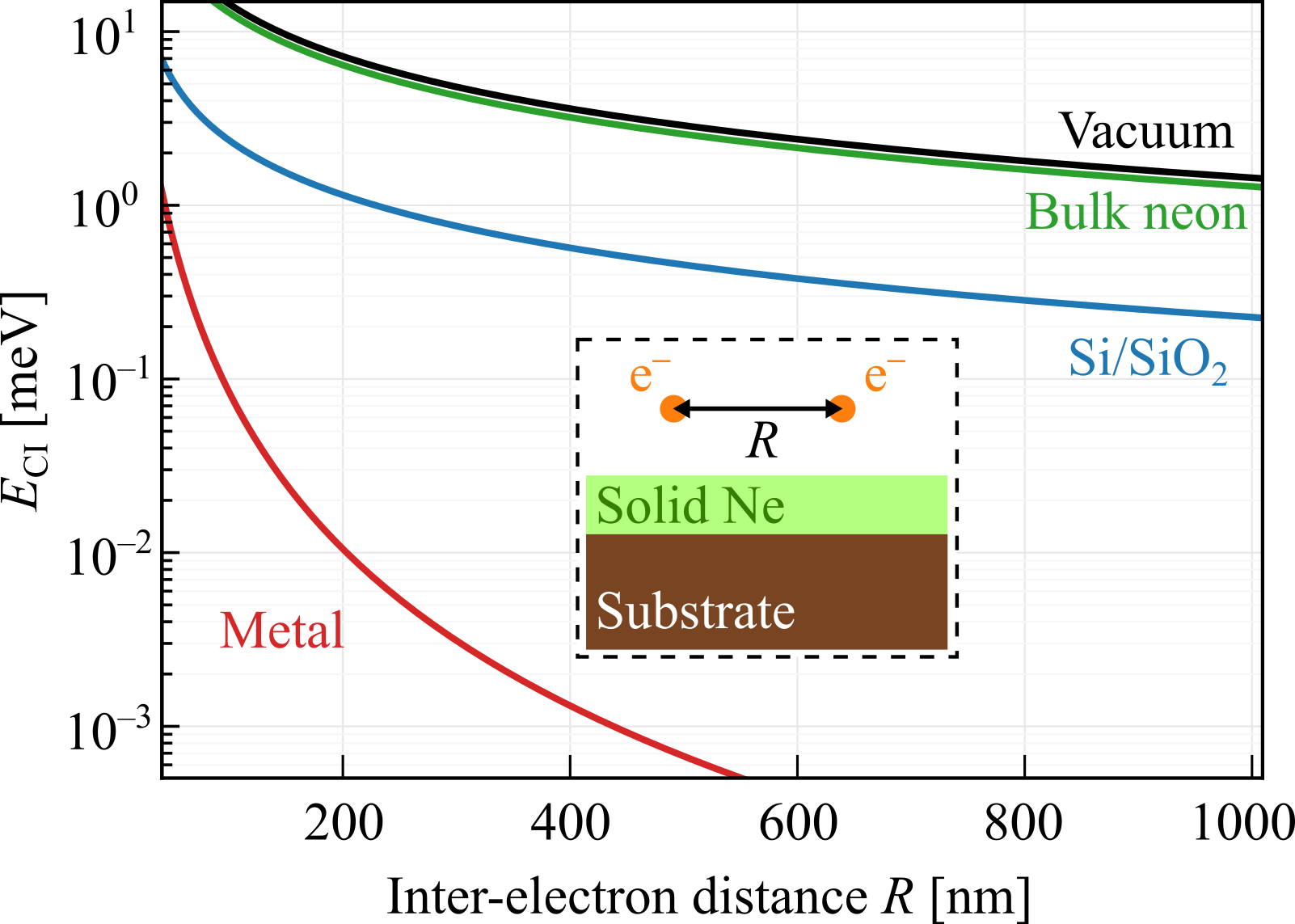}
  \caption{Coulomb interaction energy $E_\mathrm{CI}$ calculated in the point-charge approximation as a function of the inter-electron distance $R$. The black, green, blue, and red curves correspond to the vacuum, bulk neon, silicon substrate, and metallic substrate cases, respectively.}
  \label{fig:coulomb-interaction}
\end{figure}
\paragraph{Configuration.}
The electron is assumed to float in the vacuum region defined by $z>0$ in the Cartesian coordinates $(x, y, z)$. For $z<0$, three substrate configurations are considered: (i) a homogeneous solid neon substrate occupying the entire lower half-space; (ii) a multilayer stack comprising solid neon in $-t_1<z<0$, $\mathrm{SiO_2}$ in $-(t_1+t_2)<z<-t_1$, and Si for $z<-(t_1+t_2)$; and (iii) solid neon in $-t_1<z<0$, underlain by an ideal metal occupying $z<-t_1$.
The permittivities are taken as $\varepsilon_{\mathrm{Ne}}=1.244\varepsilon_0$, $\varepsilon_{\mathrm{SiO_2}}=3.9\varepsilon_0$, and $\varepsilon_{\mathrm{Si}}=11.7\varepsilon_0$.

For each configuration, we analytically derive the electrostatic Green function $\mathcal{G}(\bm{r},\bm{r}_0)$, defined by \( \phi(\bm{r}) = \int d\bm{r}_0\, \mathcal{G}(\bm{r},\bm{r}_0)\rho_e(\bm{r}_0) \), where $\phi$ is the electric potential and $\rho_e$ is the charge density~\cite{supple}.
Unless otherwise stated, we set \(t_1 = 5~\mathrm{nm}\), which corresponds to the typical neon film thickness in past experiments~\cite{zhou_single_2022}.
For the silicon substrate, we model the native oxide as a uniform $\mathrm{SiO_2}$ layer with thickness $t_2=3~\mathrm{nm}$~\cite{szekeres_ellipsometric_2013}.

\paragraph{Coulomb interaction energy.}
Figure~\ref{fig:coulomb-interaction} shows the Coulomb interaction energy between two electrons as a function of their separation $R$.
In the point-charge approximation, this energy is given by
\begin{equation}
     E_{\mathrm{CI}} (|\bm{r}_1 - \bm{r}_2|) = e^2 \mathcal{G}(\bm{r}_1,\bm{r}_2).
\end{equation}
Here $\bm{r}_j = (\bm{\rho}_j, z_g)$ is the position of the $j$-th electron in Cartesian coordinates with the mean height $z_g$ in the vertical ground state~\cite{endmatter}.
The inter-electron interactions can exhibit a strong dependence on the presence of a substrate and its dielectric constant.
Above bulk Ne, the Coulomb interaction remains above $1~\mathrm{meV}$ at a separation of $R = 1~\upmu \mathrm{m}$, a length scale comparable to the lateral extent of the resonator trapping region.
Such strong repulsion can substantially displace electrons from their trap centers and complicate the confinement of closely spaced qubits unless sufficient lateral confinement is provided.
e-Ne devices typically use thin neon films, allowing the underlying substrate to screen the inter-electron Coulomb interaction.
Above a metallic substrate, the interaction energy can fall below $0.1~\upmu\mathrm{eV}$ at $R=1~\upmu\mathrm{m}$, whereas above silicon it can remain above $0.2~\mathrm{meV}$ at the same separation.
Such repulsion makes it difficult to confine multiple electrons in close proximity unless the lateral trapping potential is sufficiently strong.

The vertical binding energy also depends on the substrate permittivity.
This dependence can produce a lateral trapping potential on the $\mathrm{meV}$ scale above metallic regions patterned on a silicon substrate~\cite{kanai_electron_2026,endmatter}.
Recent experiments demonstrate that thermal annealing and microwave driving during film growth can modify neon-film thickness and morphology~\cite{matkovic_characterizing_2025,duthaluru_real-time_2026}.
Related observations in solid argon films include changes in film growth and thickness associated with thermal gradients, as well as evaporation and regrowth following local heating~\cite{leiderer_solid_2026}.
In solid hydrogen films, morphological rearrangement has been observed below the desorption threshold and is interpreted as arising from the surface diffusion of H$_2$ molecules~\cite{conradt_solid_1994}.
For neon, microwave-induced local heating during film growth may favor thinner films near the resonator.
If such thinning occurs above the metallic regions, the enhanced substrate-induced binding could favor electron localization there~\cite{kanai_electron_2026}.

\begin{figure}[t]
  \centering
    \includegraphics[width=0.96\linewidth]{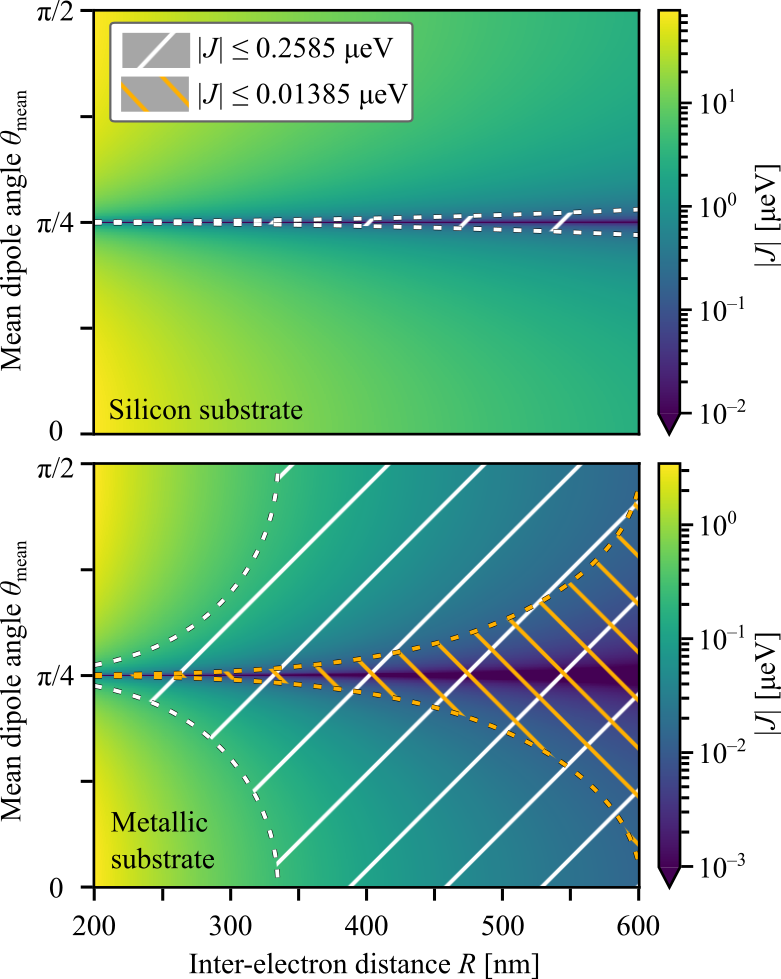}
  \caption{The magnitude of the exchange energy $|J|$ as a function of the mean angle $\theta_\textrm{mean}$ and the inter-electron distance $R$ at fixed $\theta_1 - \theta_2 = \pi/2$.
The boundaries of the white and orange dashed regions delineate the experimentally measured values of $|J|$, as described below. In the silicon substrate case, the narrow orange range is omitted.}
  \label{fig:angle_sweep}
\end{figure}

\paragraph{Two-qubit exchange coupling.}
We focus on the exchange (flip-flop) coupling $J$, which transfers a single motional excitation between the qubits.
In the two-qubit basis $\{\ket{gg},\ket{fg},\ket{gf},\ket{ff}\}$, the first and second labels refer to qubits 1 and 2, respectively, with $g$ and $f$ denoting their ground and first excited lateral motional states.
The exchange coupling is defined as \( J = \bra{fg}\hat{H}\ket{gf} \), where $\hat{H}$ is the two-electron Hamiltonian.
With the point-dipole model, the exchange coupling is approximated as
\begin{align}
    J_\mathrm{pd} &= \sum_{\alpha, \beta = {x, y, z}} d_{1}^{\alpha} \frac{\partial^2 \mathcal{G}(\bm{r}_1, \bm{r}_2)}{\partial r_{1}^{\alpha} \partial r_{2}^{\beta}} d_{2}^{\beta},
\end{align} 
where $d_j^\alpha$ is the transition-dipole component ($\alpha \in \left\{x, y, z \right\}$) of qubit $j \in \left\{1, 2 \right\}$.
For the point-dipole calculations, we use $|\bm{d}_1|=|\bm{d}_2|\simeq e\times41.538~\mathrm{nm}$, obtained by solving the single-electron Schr\"{o}dinger equation for a phenomenological single-well potential~\cite{endmatter}.

Figure~\ref{fig:angle_sweep} shows $|J|$ as a function of $R$ and $\theta_\mathrm{mean}=(\theta_1+\theta_2)/2$ at $\theta_1-\theta_2=\pi/2$.
Here, $\theta_j$ denotes the in-plane angle between the inter-electron axis and the transition dipole moment vector of qubit $j\in \{1,2\}$.
The silicon substrate only partially screens the Coulomb interaction, whereas the ideal metallic substrate cancels its leading long-range component through image charges.
Consequently, the exchange coupling decays more rapidly with electron separation above the metal, allowing appreciable coupling at short distances while suppressing coupling to more distant electrons.

Zhou~\textit{et al.}~\cite{zhou_electron_2024} spectroscopically demonstrated the simultaneous strong coupling of two e–Ne qubits to a common resonator, along with two-parameter tuning of their transition frequencies. Two-qubit entangling gates were not demonstrated in that work.
Li~\textit{et al.}~\cite{li_coherent_2025} developed a two-qubit configuration in which one qubit (bright qubit) is strongly coupled to the cavity field, whereas the other (dark qubit) displays no appreciable cavity coupling.
These two qubits are coupled via Coulomb interaction.
For the two-qubit device, they reported a negligible effective longitudinal (ZZ) interaction and an exchange coupling $|J|/h = 3.35~\mathrm{MHz}$, to which
the orange boundary in Fig.~\ref{fig:angle_sweep} corresponds.
The three-qubit value $|J|/h=62.5~\mathrm{MHz}$ is included as an additional reference scale, indicated by the white boundary.
For silicon, small $|J|$ at short separations requires restricted orientations, including for $\theta_1-\theta_2=\pi/4$.
The orange range is too narrow to display and is omitted from the silicon panel of Fig.~\ref{fig:angle_sweep}.

For the transition dipole moments assumed here, reproducing
the measured exchange coupling for electrons above silicon
requires either inter-electron separations comparable to
or larger than the lateral extent of the resonator trapping region,
or a restricted range of dipole orientations that suppress
exchange at shorter separations.
Metallic screening provides another possible explanation for weak exchange, while a location outside the region of strong cavity field could contribute to the suppressed qubit–resonator coupling of the dark qubit.
The combination of exchange, conditional, and qubit–resonator couplings therefore provides model-dependent constraints on possible trapping configurations~\cite{endmatter}.

Metallic screening may also influence qubit coherence.
For comparison, an earlier experiment~\cite{zhou_electron_2024} using a device with a dedicated trap electrode reported $T_1 \simeq 48.2~\upmu\mathrm{s}$ and Ramsey coherence time $T_2^* \simeq 42.8~\upmu\mathrm{s}$ for a qubit at its charge sweet spot.
A single refocusing $\pi$ pulse extended its Hahn-echo coherence time to $T_2^{\mathrm{echo}} \simeq 92.9~\upmu\mathrm{s}$.
In a subsequent study without that dedicated central trap electrode~\cite{li_solid_2026}, Ramsey coherence times were
$T_2^* \simeq 3.5$--$8.2~\upmu\mathrm{s}$ for three
electron-on-neon qubits, each measured at its charge sweet spot.
Theoretical estimates predict intrinsic phonon-limited coherence times well beyond the experimentally observed
$0.1~\mathrm{ms}$ scale at $T\sim10~\mathrm{mK}$~\cite{zou_electron_2026}.
Therefore, screening of electric-field fluctuations from surrounding charge sources by the dedicated trap electrode may be one factor contributing to the longer coherence time reported in the earlier device.

\begin{figure}[t]
  \centering
    \includegraphics[width=0.85\linewidth]{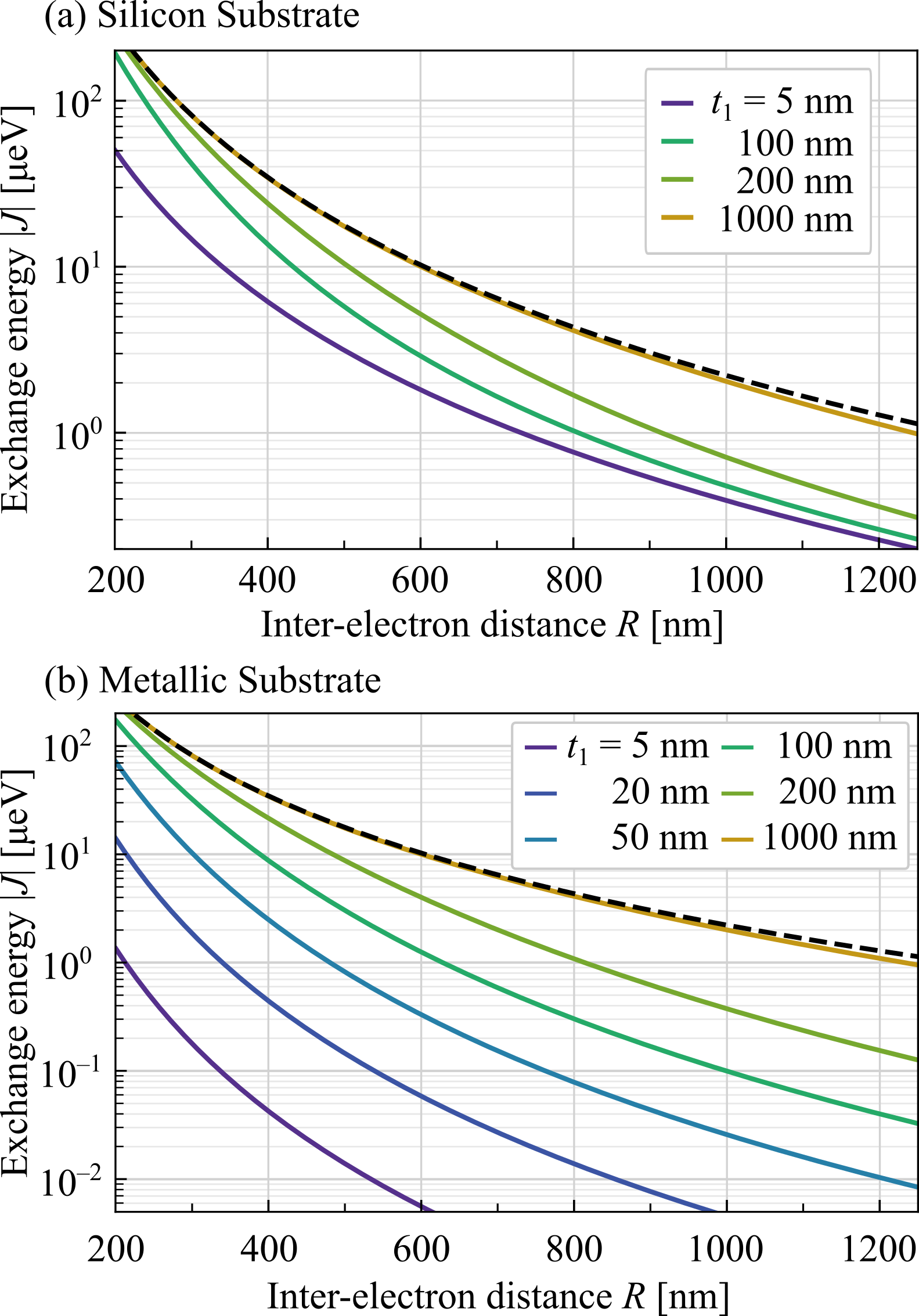}
    \caption{The magnitude of the exchange energy $|J|$ calculated by the point-dipole model for $\theta_1=\theta_2=\pi/2$ and several neon-film thicknesses $t_1$ above the silicon substrate (a) and the metallic substrate (b).
    The black dashed line corresponds to the bulk neon case.}
    \label{fig:thickness_J}
\end{figure}
\paragraph{Neon-Thickness Dependence.}
Neon films can exhibit spatial variations in thickness~\cite{matkovic_characterizing_2025}, and electrons may therefore be trapped above regions thicker than the $5~\mathrm{nm}$ film assumed here.
This value corresponds to the lower end of the $5$--$10~\mathrm{nm}$ range~\cite{zhou_single_2022}.
At fixed electron separation and transition dipole moments, increasing the neon-film thickness weakens substrate screening and strengthens the inter-electron Coulomb interaction.
Away from configurations involving angular cancellation, this is expected to enhance $|J|$ as well, although its detailed thickness dependence depends on the dipole orientations.
For orientations away from angular cancellation, thicker local films are therefore not expected to account for the small exchange coupling reported by Li~\textit{et al.}~\cite{li_coherent_2025}.

As a reference, Fig.~\ref{fig:thickness_J} shows the magnitude of the exchange energy $|J|$ for $\theta_1=\theta_2=\pi/2$ with various neon thicknesses $t_1$.
These results suggest that as the solid neon layer becomes thicker, a longer distance is needed to suppress crosstalk.
For micrometer-thick neon films, the exchange energy approaches the bulk-neon limit, potentially compromising the performance of our proposed compact gate design.

\paragraph{Compact multiqubit architecture.}
\begin{figure}[t]
  \centering
    \includegraphics[width=0.8\linewidth]{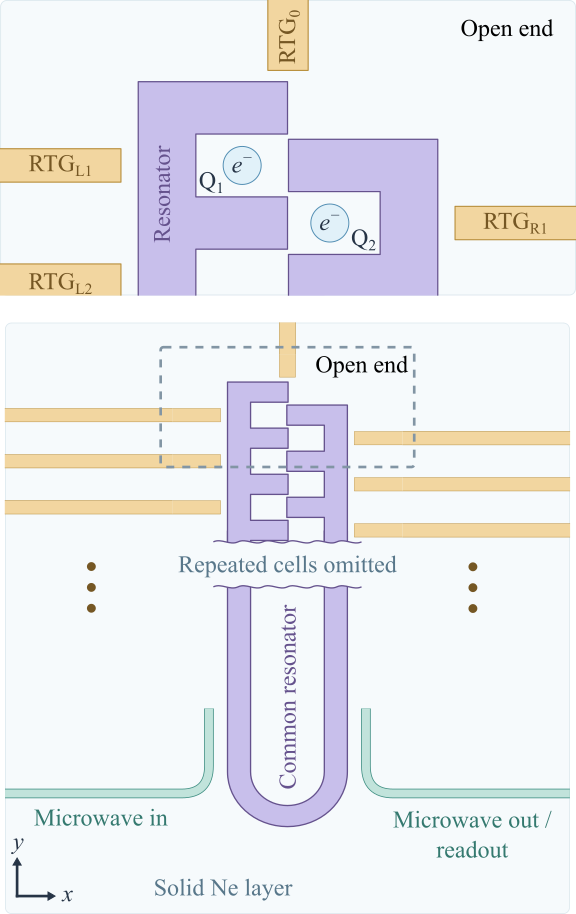}
  \caption{Proposed multiqubit resonator unit with metallic screening and local gate control. Top: Detail of two trapping regions for qubits \(Q_1\) and \(Q_2\). Bottom: Extended layout with repeated, staggered trapping regions near the open ends. The dashed box marks the region enlarged above. Purple, gold, and green denote the superconducting resonator, independently biased resonator trap gates (RTGs), and microwave coupling lines, respectively.}
  \label{fig:two-qubit-gate}
\end{figure}
On the basis of these results, we propose the compact multiqubit resonator unit shown in Fig.~\ref{fig:two-qubit-gate} as an extension of existing e–Ne devices.
The proposed design retains the superconducting resonator for microwave control and dispersive readout and incorporates a repeating array of staggered trapping regions with independently biased local trap gates.
The layout is intended to accommodate several qubits near the resonator voltage antinode.
Metallic screening, site spacing, and local electrostatic tuning would then be used together to retain selected direct couplings while suppressing unwanted interactions.
Detuning neighboring qubits such that $|\omega_i-\omega_{i+1}|\gg |J|/\hbar$, where $\omega_i$ is the transition angular frequency of qubit $i$, suppresses resonant excitation exchange.
Residual conditional interactions during idle periods and charge-noise-induced dephasing at detuned operating points must both be assessed along the control trajectory.
Repeated, staggered projections of the two resonator arms are intended to define separate trapping regions, each hosting a single electron charge qubit. The arms are joined at a rounded closed end, while the trapping array is confined to the open-end section near the voltage antinode of the intended differential mode.

Spatially distributed, independently biased RTGs provide multiple parameters for adjusting the confinement at each site. Joint optimization of these voltages is intended to tune both the qubit transition frequencies and the overlap of their transition dipoles with the resonator field, enabling the selection of qubits with strong qubit--resonator coupling.
Spectral detuning provides an additional means of reducing the participation of spectator qubits in a selected operation. The proposed control thus extends the demonstrated two-parameter tuning of two qubits~\cite{zhou_electron_2024} to an array of separately controlled trapping regions. 
The multiple electrostatic control parameters could enable selective two-qubit gates between any chosen pair of nearby qubits.
Additional trap electrodes, not shown in Fig.~\ref{fig:two-qubit-gate}, could be incorporated beneath the solid-neon film at the intended electron trapping sites. These electrodes would provide further control over local confinement and could enhance metallic screening of direct Coulomb interactions between non-neighboring electrons.
The attainable site density, coupling strengths, and degree of local control in this patterned geometry require electrostatic and electromagnetic modeling.
A central design objective is to bring the selected qubits into mutual resonance while keeping both near operating points with low charge-noise sensitivity.

We propose varying the local gate biases and measuring the exchange and conditional couplings, qubit transition frequencies, and qubit--resonator couplings.
Comparing their bias dependence with predictions from candidate trapping models could constrain the electron positions and lateral confinement.

As an illustrative benchmark, we adopt
$|J_\mathrm{exp}|/h=3.35~\mathrm{MHz}$, reported for
the two-qubit device~\cite{li_coherent_2025}.
Ideal resonant exchange implements an iSWAP in $t_g=\pi\hbar/(2|J_{\mathrm{exp}}|)\simeq74.6~\mathrm{ns}$, up to calibrated single-qubit phases.
For this estimate, we assume that both qubits retain the reported single-qubit sweet-spot relaxation and Ramsey times, $T_1=48.2~\upmu\mathrm{s}$ and $T_2^*=42.8~\upmu\mathrm{s}$~\cite{zhou_electron_2024}, throughout the gate.
Approximating Ramsey decay as exponential gives $\Gamma_1=1/T_1$ and $\Gamma_\phi=1/T_2^*-1/(2T_1)$ for both qubits.
For independent, weak Markovian relaxation and dephasing,
the leading-order average gate infidelity is~\cite{abad_universal_2022}
\begin{equation}
1-F_{\mathrm{avg}}^{(\mathrm{diss})}
\simeq\frac{2t_g}{5}\sum_{i=1}^{2}
(\Gamma_{1,i}+\Gamma_{\phi,i}),
\end{equation}
where $F_{\mathrm{avg}}^{(\mathrm{diss})}$ denotes the average two-qubit gate fidelity, accounting only for these dissipative processes, with the ideal iSWAP gate as the target operation.
This gives $F_{\mathrm{avg}}^{(\mathrm{diss})}\simeq99.80\%$.
The gate time used in this estimate corresponds to ideal
resonant exchange; the microwave-driven cross-resonance and bSWAP gates in Ref.~\cite{li_coherent_2025} follow different protocols.
The adopted $T_1$ already includes Purcell decay in the
reference device.
This estimate neglects switching overhead, residual detuning, unwanted conditional phase, and leakage. 
Realizing this benchmark requires validating the assumed
coupling strength and coherence times for both qubits
in the proposed geometry.

\paragraph{Conclusion}
We have quantified the substrate screening of Coulomb interactions on thin solid neon films.
Two-qubit architectures proposed for e--He systems~\cite{kawakami_blueprint_2023,
beysengulov_coulomb_2024,leinonen_design_2026} can also be adapted to e--Ne systems, but exploiting the strong screening identified here offers a route to more compact designs.
The multiqubit resonator architecture in Fig.~\ref{fig:two-qubit-gate} illustrates this approach, combining screening with local gate control.

For a millimeter-scale resonator, an assumed usable trapping length of $250~\upmu\mathrm{m}$ near an open end and a site pitch of $300~\mathrm{nm}$ give a geometrical capacity of approximately 830 trapping sites per resonator.
The achievable number of controllable qubits will depend on maintaining confinement, coherence, and selective coupling across the array.
This estimate nevertheless illustrates the potential for dense integration within a single resonator.
For comparison, representative experiments with single ion chains have demonstrated gate-based quantum simulation using 32 computational qubits and four auxiliary qubits~\cite{chernyshev_pathfinding_2026}, analog quantum simulation with 51 ions~\cite{kraft_bounded_2026}, and multiqubit quantum memory in a 218-ion
chain~\cite{yao_experimental_2022}.
Connecting multiple resonator units through engineered inter-unit couplings could further extend this approach to two-dimensional arrays of electron-on-neon qubits.

\begin{acknowledgments}
\emph{Acknowledgments:}
The authors acknowledge Kaiwen~Zheng, Dafei~Jin, Paul~Leiderer, and David~Rees for the fruitful discussion.
This work is supported by the Air Force Office of Scientific Research under Grant No.~FA9550-26-1-B212 and the National Science Foundation under Grant No.~PHY-2607484.
\end{acknowledgments}

\paragraph{AI Use Disclosure.}
We used ChatGPT (GPT-6 Astra) to assist with code development.
The authors selected the numerical methods, reviewed the code,
and verified the results.

\clearpage
\appendix

\section{End Matter}
\paragraph{Schr\"{o}dinger model and Hartree approximation.}
We describe the two-electron system by the wavefunction $\Psi(\bm{r}_1,\bm{r}_2)$.
The Hamiltonian $\hat{H}$ includes potential energy contributions from the neon barrier $V_{\mathrm{ne}}$, electrostatic interactions $V_{\mathrm{el}}$, and lateral confinement $V_{\mathrm{tr}}$.
The neon barrier is $V_{\mathrm{ne}}(z)=0.7~\mathrm{eV}$ for $z<0$ and zero for $z>0$.
The total electrostatic interaction potential is defined by \( V_{\mathrm{el}} = \sum_j V_\mathrm{self} (z_j) + e^2 \mathcal{G} (\bm{r}_1, \bm{r}_2)\) with $V_{\mathrm{self}} (z) = e^2 \mathcal{G}_\mathrm{ind} (\bm{r}, \bm{r}) / 2$, where the Green's function is decomposed as $\mathcal{G} = \mathcal{G}_0 + \mathcal{G}_\mathrm{ind}$ with the bare Coulomb Green function $\mathcal{G}_{0}$ and the contribution induced by the dielectric interfaces $\mathcal{G}_\mathrm{ind}$~\cite{supple}.
$V_\mathrm{tr}$ is the lateral trapping potential and consists of two trapping potentials $V_{\mathrm{tr}}^{(j)}(\bm{\rho}_j)$ with $j\in\{1, 2\}$.
Following the phenomenological model by Zhou~\textit{et al.}~\cite{zhou_single_2022}, we suppose that the $j$-th electron is trapped by a single-well trapping potential:
\begin{eqnarray}
    V_{\mathrm{tr}}^{(j)}(\bm{\rho}_j) &=& \frac{1}{2} m_e \omega^2_\parallel \left( s_j^2+\beta s_j^4 \right) + \frac{1}{2} m_e \omega_\perp^2 u_j^2.
    \label{eq:trap-potential}
\end{eqnarray}
Here $s_j$ and $u_j$ denote coordinates along the soft axis of trap $j$ and the perpendicular in-plane axis, respectively.
The parameters are
$k_\parallel = m_e \omega_\parallel^2 = 5.536~\mathrm{meV / \upmu m^2}$ and $\beta = 15.4~\upmu \mathrm{m^{-2}}$.
To strongly confine the electron along the $u_j$-axis, we take $\omega_\perp = 3 \omega_\parallel$.
With a single electron trapped by this trapping potential, the excitation energy along the soft axis is $\Delta E_\parallel \approx 22.071~\upmu \mathrm{eV}$, and the transition dipole moment along it is $d_\mathrm{tra} \approx e\times 41.538~\mathrm{nm}$, which nearly saturates the upper bound governed by the f-sum rule:
\( \lvert d_\mathrm{tra} \rvert \le {e \hbar}/{\sqrt{2m_e \Delta E_\parallel}} \approx e \times 41.549~\mathrm{nm} \).

\begin{table}[h]
\centering
\caption{Calculated vertical confinement of an electron above
solid neon for the three substrate configurations.
$z_g$ is the mean height above
the neon surface in the ground vertical state, $E_g^{(z)}$ is the ground-state energy relative to the vacuum level, and $\Delta E^{(z)}$ is the vertical excitation energy.}
\label{tab:quantum-z}
\begin{tabular}{|c|| c | c | c |}
 \hline
$\text{Substrate}$ & $z_g~[\mathrm{nm}]$ & $E^{(z)}_g~[\mathrm{meV}]$ & $\Delta E^{(z)}~[\mathrm{meV}]$ \\ [0.5ex] 
 \hline\hline
 \text{Bulk Ne} & 1.88 & $-15.69$ & $12.59$ \\ 
 \hline
 \text{Silicon} & 1.49 & $-53.27$ & $26.65$ \\
 \hline
 \text{Metal} & 1.39 & $-67.33$ & $31.97$ \\ 
 \hline
\end{tabular}
\end{table}
The binding and excitation energies in the $z$-direction are much larger than the lateral excitation energies considered here, and the vertical quantum state is assumed to be frozen in the ground state $\chi_g$: \( \Psi (\bm{r}_1, \bm{r}_2) = \Phi(\bm{\rho}_1, \bm{\rho}_2) \chi_g(z_1) \chi_g(z_2) \)
with $\bm{\rho}=(x, y)$.
The vertical wavefunction $\chi_n$ is obtained by solving the Schr\"{o}dinger equation in the z-direction:
\begin{align}
     E_n^{(z)} \chi_n (z) &= \left[-\frac{\hbar^2}{2m_e} \partial_{z}^2 + V_{\mathrm{self}} (z) + V_\mathrm{ne} (z) \right] \chi_n(z)
\end{align}
with a potential cutoff at $z = 0.23~\mathrm{nm}$~\cite{zhou_single_2022, cole_image-potential-induced_1969, cole_electronic_1971}.
Here $n$ labels the vertical eigenstates, with $n=g$ denoting the ground state.
Table~\ref{tab:quantum-z} summarizes the relevant quantities.

We further neglect inter-site orbital overlap and construct state-specific Hartree product states~\cite{beysengulov_coulomb_2024}:
\begin{equation}
\Phi_A(\bm{\rho}_1, \bm{\rho}_2) \approx \Phi^{(H)}_A(\bm{\rho}_1, \bm{\rho}_2) := \phi_{A}^{(1)}(\bm{\rho}_1) \phi_{A}^{(2)}(\bm{\rho}_2)
\end{equation}
for a given two-electron configuration 
$A \equiv (\alpha,\beta)$, where \(\alpha\) labels the orbital state of electron/site 1 and \(\beta\) labels the orbital state of electron/site 2.
The ground state and first excited state are denoted as $g$ and $f$, respectively.
With both electrons restricted to the vertical ground state, the state-specific Hartree equations for $\phi_{A}^{(j)} (\bm{\rho}_j)$ with the site index $j\in\{1, 2\}$ can be written as
\begin{align}
&\epsilon_{A}^{(j)} \phi_{A}^{(j)} = \left[ -\frac{\hbar^2}{2m_{\mathrm{e}}} \nabla_{\rho_j}^2 + V_\mathrm{tr}^{(j)} (\bm{\rho}_j) \right. \nonumber \\
& \hspace{45pt} + \iint dz_j\,dz_l\, |\chi_g(z_j)|^2 |\chi_g(z_l)|^2 \nonumber\\
& \hspace{50pt} \left. \times \int\,d\bm{\rho}_l\, V_\mathrm{e-e}(\bm{r}_j, \bm{r}_l) \left|\phi_{A}^{(l)}(\bm{\rho}_{l})\right|^2 \right] \phi_{A}^{(j)}
\end{align}
with $l=3-j$.
Here $\epsilon_A^{(j)}$ are the self-consistent Hartree orbital energies, and $V_{\mathrm{e-e}}(\bm{r}_1,\bm{r}_2)=e^2\mathcal{G}(\bm{r}_1,\bm{r}_2)$ is the inter-electron interaction.

\begin{figure}[b]
  \centering
    \includegraphics[width=0.83\linewidth]{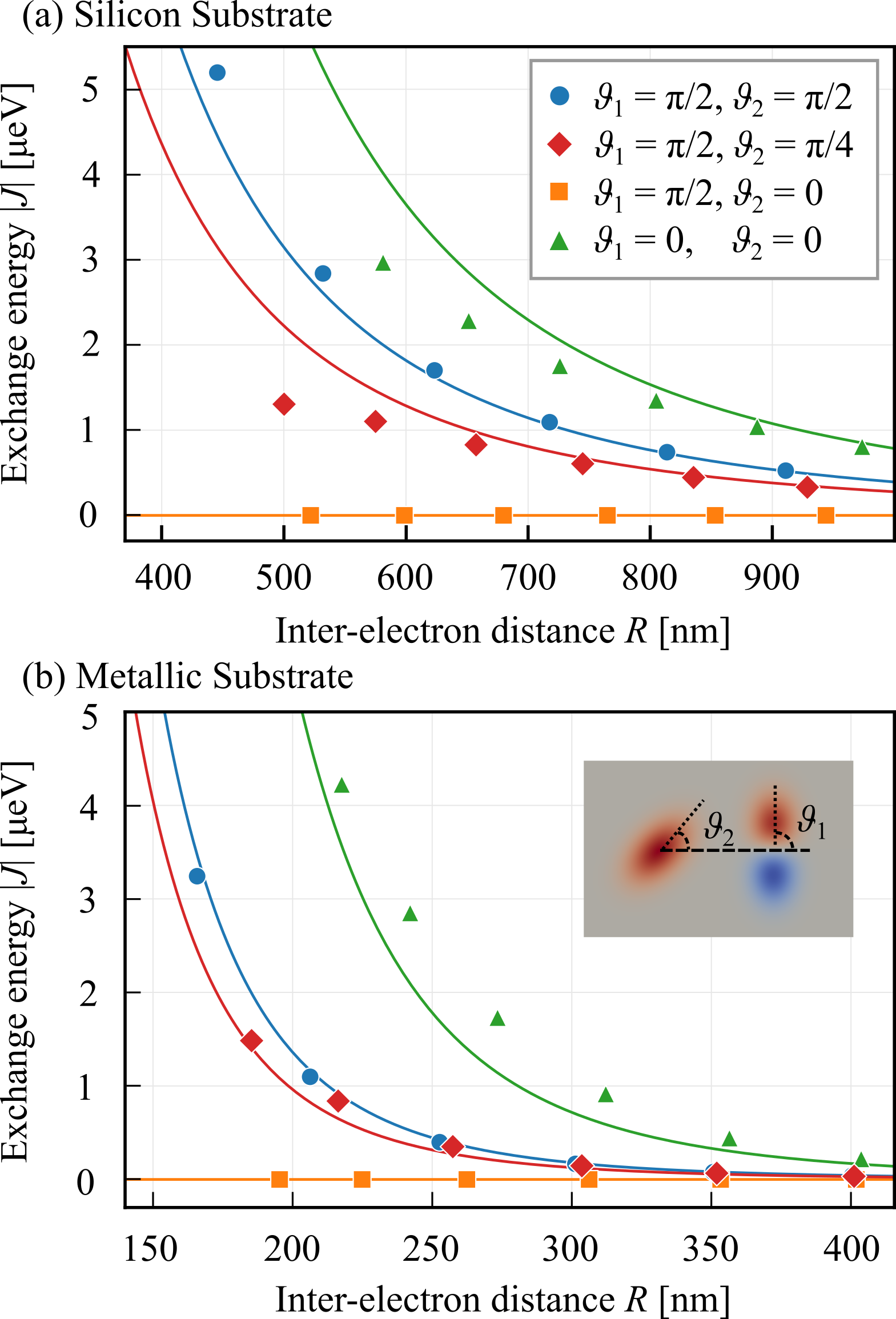}
    \caption{The magnitude of the exchange energy $|J|$ as a function of the inter-electron separation $R$ with the silicon substrate (a) or the metallic substrate (b).
    Solid curves and discrete markers denote the point-dipole model and the wavefunction-based calculation, respectively.}
    \label{fig:inter-electron_J}
\end{figure}

\paragraph{Point-dipole model.}
To assess the validity of the point-dipole approximation, we compare its predictions for $J$ with those from the self-consistent Hartree calculations in Fig.~\ref{fig:inter-electron_J}.
We analyze four configurations of the soft-axis orientations; each configuration is specified by the two angles $\vartheta_1$ and $\vartheta_2$, defined as the angles between the trap-center axis (the line connecting the two trap centers) and the soft axes of traps 1 and 2, respectively.
We define the mean in-plane electron separation as
\( R := \int d^2\bm{\rho}_1\,d^2\bm{\rho}_2\, |\bm{\rho}_1-\bm{\rho}_2|\, |\Phi^{(H)}_{gg}(\bm{\rho}_1, \bm{\rho}_2) |^2 \).

The state-specific Hartree configurations $\ket{A}_H$ are generally nonorthogonal.
We define the Hamiltonian and overlap matrices as $H_{AB}={}_H\!\bra{A}\hat{H}\ket{B}_H$ and $S_{AB}={}_H\!\braket{A|B}_H$, respectively, and apply L\"{o}wdin symmetric orthogonalization~\cite{lowdin_nonorthogonality_1950}.
The resulting orthonormal diabatic basis is defined by \(\ket{\widetilde{A}} = \sum_B \ket{B}_H (S^{-1/2})_{BA}\) with \(A,B\in\{gg,fg,gf,ff\}\), where \(S\) is the overlap matrix of the Hartree configurations.
The Hamiltonian in this orthonormal diabatic basis is
$H_{\mathrm{orth}} = S^{-1/2} H S^{-1/2}$.
Provided that no overlap eigenvectors are discarded in the orthogonalization, the effective exchange coupling is 
$J = (H_{\mathrm{orth}})_{fg,gf}$.

The point-dipole model captures the overall distance dependence of the exchange coupling, although noticeable configuration-dependent deviations from the Hartree results remain.
The mean electron separation generally differs from the separation between the trap centers.

\begin{figure}[b]
  \centering
    \includegraphics[width=0.83\linewidth]{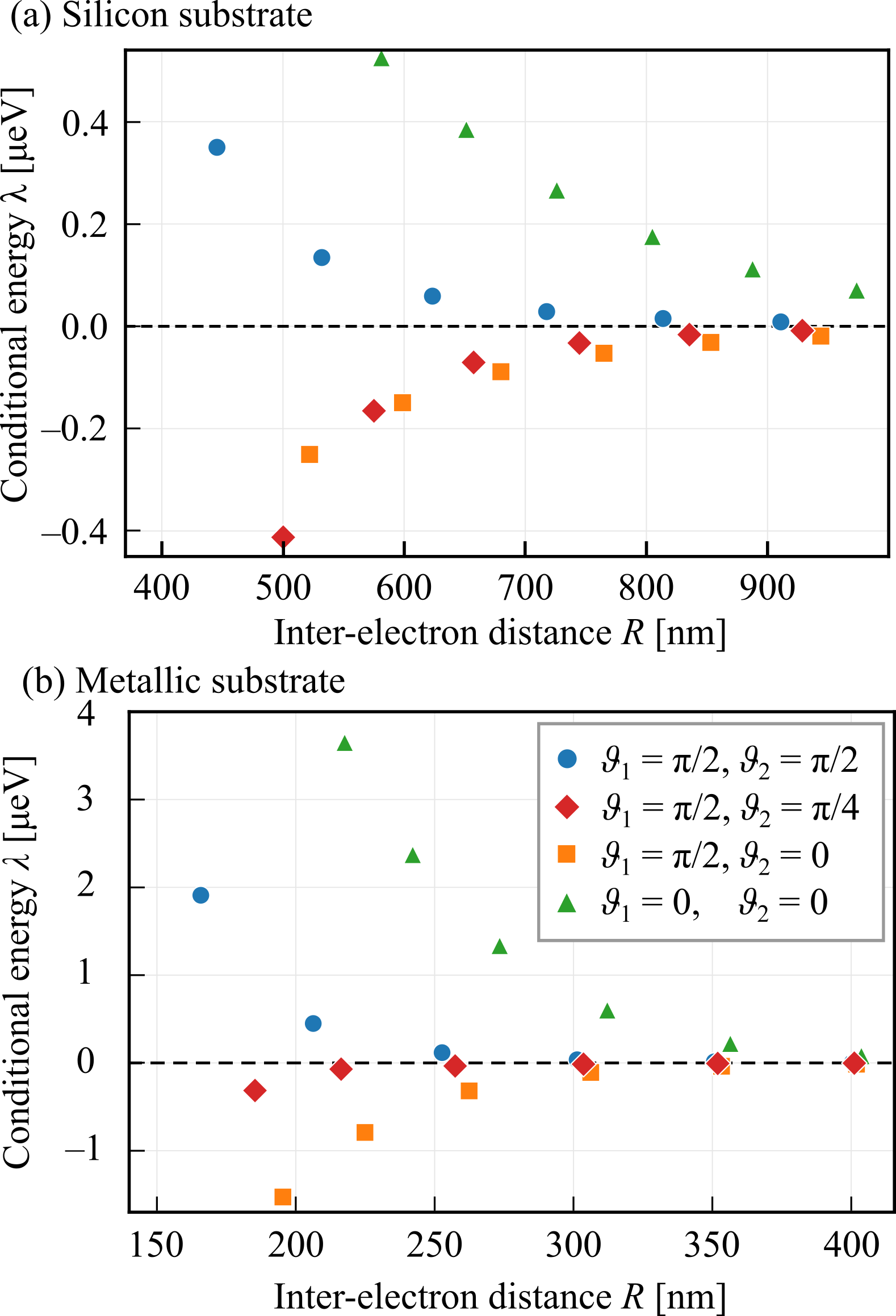}
    \caption{The diagonal conditional interaction energy $\lambda$ as a function of the inter-electron separation $R$ with the silicon substrate (a) or the metallic substrate (b).}
    \label{fig:inter-electron_lambda}
\end{figure}
\paragraph{Diagonal conditional interaction.}
The diagonal conditional interaction is defined in the orthonormal diabatic basis as
\( \lambda = (H_{\mathrm{orth}})_{ff,ff} -(H_{\mathrm{orth}})_{fg,fg} -(H_{\mathrm{orth}})_{gf,gf} +(H_{\mathrm{orth}})_{gg,gg} \).
The corresponding longitudinal interaction term is
\begin{equation}
    H_{ZZ}=\frac{\lambda}{4}\sigma_z^{(1)}\sigma_z^{(2)},
\end{equation}
where $\sigma_z^{(j)}$ is the Pauli-$z$ operator acting on qubit $j$ in the effective $\{g,f\}$ state space.
Figure~\ref{fig:inter-electron_lambda} shows $\lambda$ for the same four configurations as in Fig.~\ref{fig:inter-electron_J}.
The sign and magnitude of $\lambda$ depend strongly on the trap orientations. In particular, the orthogonal soft-axis configuration retains a finite conditional interaction even when the exchange coupling is strongly suppressed.
A negligible effective longitudinal coupling was inferred from spectroscopy in Ref.~\cite{li_coherent_2025}.

\paragraph{Small exchange interaction in the experiments.}
A strongly reduced transition dipole provides another possible explanation for the small exchange coupling observed in the experiments~\cite{li_coherent_2025}.
Such a reduction is difficult to obtain simply by tightening a nearly harmonic single-well confinement while retaining the observed transition frequency.
An asymmetric double well can, however, support a small transition dipole at the same excitation energy~\cite{stockklauser_strong_2017}.
Localization of the ground and first excited states in opposite wells produces a state-dependent charge displacement that can enhance the conditional interaction, depending on the geometry and the charge response of the other qubit.
A sufficiently short separation between the wells can reduce this displacement.
To estimate the relevant length scale, we use the reported equivalent differential-mode impedance \(Z = 598.5~\Omega\) and the resonator frequency \(\omega_r/2\pi=5.668~\mathrm{GHz}\)~\cite{li_solid_2026}.
These give a zero-point mode voltage \(V_{\mathrm{zpf}}=\omega_r\sqrt{\hbar Z/2}\simeq6.3~\upmu\mathrm{V}\)~\cite{blais_circuit_2021}.
Assuming a local field-to-mode-voltage ratio of at most $1~\upmu\mathrm{m}^{-1}$, we obtain the zero-point electric-field amplitude $E_{\mathrm{zpf}}\lesssim6.3~\mathrm{V/m}$.
Within the electric-dipole approximation, the measured coupling \(g_b/2\pi = 3.76~\mathrm{MHz}\) between the bright qubit and the resonator~\cite{li_coherent_2025} then implies \(|d_{b,\parallel}|\gtrsim e\times 2.5~\mathrm{nm}\), where $d_{b,\parallel}$ is the component of the transition dipole moment along the local mode field.
In a localized two-state double-well model, this implies an interwell separation of at least approximately $5~\mathrm{nm}$.
If the same trapping mechanism produces comparable interwell separations for both qubits, explanations of the weak conditional interaction inferred from spectroscopy based on closely spaced wells must be assessed quantitatively against this length scale.

\clearpage
\onecolumngrid

\begin{center}
{\large\bfseries Supplemental Material: Derivation of the Electrostatic Green Functions}
\end{center}

We consider planar layers parallel to the $xy$ plane, with the neon–vacuum interface at $z=0$. Vacuum (region $0$) occupies $z>0$, and solid neon (region $\mathrm{Ne}$) occupies $-t_1<z<0$. For the silicon substrate, an $\mathrm{SiO_2}$ layer (region $1$) of thickness $t_2$ occupies $-(t_1+t_2)<z<-t_1$, and a semi-infinite silicon substrate (region $2$) occupies $z<-(t_1+t_2)$. The relative permittivity of region $a$ is $\kappa_a=\varepsilon_a/\varepsilon_0$, with $\kappa_0=1$. For the metallic substrate, the oxide and silicon regions are replaced by an ideal metal occupying $z<-t_1$. For bulk neon, solid neon occupies the entire region $z<0$.

The electrostatic Green function satisfies the Poisson equation:
\begin{eqnarray}
    \nabla \cdot [\varepsilon(z) \nabla \mathcal{G}(\bm{r}, \bm{r}')] = -\delta(\bm{r} - \bm{r}').
\end{eqnarray}
For a fixed source point $\bm{r}'$, we require $\mathcal{G}(\bm{r},\bm{r}')\to 0$ as $|\bm{r}|\to\infty$ within the nonmetallic domain.
At each dielectric interface without free surface charge, both $\mathcal{G}$ and $\varepsilon(z)\partial_z\mathcal{G}$ are continuous.
For the ideal metallic substrate held at fixed potential, we impose the homogeneous Dirichlet condition $\mathcal{G}(\bm{r},\bm{r}')=0$ at $z=-t_1$.
Using translational invariance and rotational symmetry in the
$xy$ plane, we introduce the two-dimensional Fourier transform
\begin{align}
\tilde{\mathcal{G}}(k,z,z')
&= \int d^2\rho\,
e^{-i\bm{k}\cdot(\bm{\rho}-\bm{\rho}')}
\mathcal{G}(\bm{r},\bm{r}'),
\end{align}
where $\bm{r}=(\bm{\rho},z)$ and $\bm{\rho}=(x,y)$,
and $k=|\bm{k}|$ is the magnitude of the in-plane wave vector.
The transformed Green function satisfies
\begin{eqnarray}
\frac{d}{dz} \left[ \varepsilon(z) \frac{d \tilde{\mathcal{G}}}{dz} \right] - \varepsilon(z) k^2 \tilde{\mathcal{G}} = -\delta(z - z').
\end{eqnarray}

For $z,z'>0$, we write $\mathcal{G}=\mathcal{G}_0+\mathcal{G}_{\mathrm{ind}}$, where $\mathcal{G}_0$ is the bare Coulomb Green function and $\mathcal{G}_{\mathrm{ind}}$ is the contribution induced by the interfaces. These terms are given by
\begin{align}
    \mathcal{G}_{0} (\bm{r}, \bm{r}') &= \frac{1}{4\pi \varepsilon_0} \int_0^\infty dk\, J_0(k |\bm{\rho} - \bm{\rho}'|) e^{-k|z - z'|} \nonumber \\
    &= \frac{1}{4\pi \varepsilon_0} \frac{1}{\sqrt{|\bm{\rho} - \bm{\rho}'|^2 + |z - z'|^2}} \\
    \mathcal{G}_\mathrm{ind} (\bm{r}, \bm{r}') &= 
    \frac{1}{4\pi \varepsilon_0}\int_0^\infty dk\, \Lambda_\mathrm{ind}(k) J_0(k |\bm{\rho} - \bm{\rho}'|) e^{-k(z + z')}
\end{align}
with the reflection coefficients:
\begin{align}
\Lambda_\mathrm{ind} = \frac{\alpha_{0, \mathrm{Ne}} + \Lambda_\mathrm{sub} e^{-2kt_1}}{1 + \alpha_{0, \mathrm{Ne}} \Lambda_\mathrm{sub} e^{-2kt_1}},\hspace{10pt} 
\Lambda_\mathrm{sub} = \frac{\alpha_{\mathrm{Ne}, 1} + \alpha_{1, 2} e^{-2kt_2}}{1 + \alpha_{\mathrm{Ne}, 1} \alpha_{1, 2} e^{-2kt_2}}
\end{align}
where \(\alpha_{a, b} = (\kappa_a - \kappa_b)/(\kappa_a + \kappa_b) \), and $J_0$ denotes the Bessel function of the first kind of order zero.
The expression for $\Lambda_{\mathrm{sub}}$ applies to the silicon stack. For an ideal metal directly beneath the neon film, we set $\Lambda_{\mathrm{sub}}=-1$.
For bulk neon, $\Lambda_{\mathrm{ind}}=\alpha_{0,\mathrm{Ne}}$.

In other domains, the Green function is  
\begin{align}
    \mathcal{G} (\bm{r}, \bm{r}') &= \frac{1}{4 \pi \varepsilon_0} \int_0^\infty dk\, J_0(k |\bm{\rho} - \bm{\rho}'|) \frac{2}{1 + \kappa_\mathrm{Ne}} \frac{e^{kz} + \Lambda_\mathrm{sub} e^{-2kt_1} e^{-kz}}{1 +\alpha_{0,\mathrm{Ne}} \Lambda_\mathrm{sub} e^{-2kt_1}} e^{-kz'}
\end{align}
in $-t_1<z<0$ and $0 < z'$, and
\begin{align}
\mathcal{G} (\bm{r}, \bm{r}') = 
\begin{dcases}
\frac{1}{4 \pi\varepsilon_0} \int_0^\infty dk\, J_0(k |\bm{\rho} - \bm{\rho}'|) \frac{2e^{-kz}}{1 + \kappa_\mathrm{Ne}} \frac{e^{kz'} + \Lambda_\mathrm{sub} e^{-2kt_1} e^{-kz'}}{1 + \alpha_{0, \mathrm{Ne}} \Lambda_\mathrm{sub} e^{-2kt_1}} & 0 < z \\
\frac{1}{4 \pi \varepsilon_0} \int_0^\infty dk\, J_0(k |\bm{\rho} - \bm{\rho}'|) \frac{-\alpha_{0, \mathrm{Ne}} e^{k z_{\max}} + e^{-k z_{\max}}}{\kappa_\mathrm{Ne}} \frac{e^{k z_{\min}} + \Lambda_\mathrm{sub} e^{-2kt_1} e^{-k z_{\min}}}{1 + \alpha_{0, \mathrm{Ne}} \Lambda_\mathrm{sub} e^{-2kt_1}}. & -t_1 < z < 0 \\
\end{dcases}
\end{align}
with $z_{\min} = \min(z, z')$ and $z_{\max} = \max (z, z')$ in $-t_1 < z' < 0$.

\end{document}